# Acoustooptic Interaction in Barium Beta-Borate Crystals

Martynyuk-Lototska I., Dudok T., Dyachok Ya., Burak Ya. and Vlokh R.

Institute of Physical Optics, 23 Dragomanov Str., 79005 L'viv, Ukraine

## Abstract

Basing on the measurements of photoelastic coefficients, acoustic waves velocities and the refractive indices, the possible geometries for acoustooptic interaction in $\beta$-BaB$_2$O$_4$ crystals have been analysed. It has been shown that the acoustooptic figure of merit achieves its maximum value when the transverse acoustic wave $v_{13}$ with the velocity 817m/s is propagated in YZ-plane. In this case $M_2$=95.35×10$^{-15}$s$^3$/kg.

It has been shown in our previous paper [1] that the acoustooptic figure of merit (AOFM) for $\alpha$-BaB$_2$O$_4$ and Li$_2$B$_4$O$_7$ crystals can be as high as $M_2$=243.4×10$^{-15}$s$^3$/kg and $M_2$=2.57×10$^{-15}$s$^3$/kg in case of the optimised geometry for acoustooptic (AO) interaction. Such the magnitudes of AOFM appear due to AO interaction with the slowest acoustic waves. The direction of propagation of these waves does not coincide with the principle (crystallophysical) directions but is determined by anisotropy of the ultrasonic wave velocities. The crystals of $\beta$-BaB$_2$O$_4$ exhibit rather high values of AOFM, too [2]. Therefore the goal of the present paper is to analyse all the possible geometries for the AO interactions in $\beta$-BaB$_2$O$_4$ crystals on the basis of construction of indicative surfaces for the acoustic wave velocities.

The measured velocities of the longitudinal and transverse ultrasonic waves in $\beta$-BaB$_2$O$_4$ crystals and the possible geometries satisfying the Bragg condition are collected in Table 1.

Using the experimental data for the ultrasonic velocity, piezooptic coefficients and the refractive indices, the AOFM $M_2=p^2_{ef}n^6/\rho v^3$ may be calculated for the cases when the optical beam and the acoustic wave propagate along the principal directions (see Table 1). Let us take into account that the measured value of the velocity for one of the slowest transverse acoustic wave manifesting bearable attenuation is $v_{13}$= 880 m/s and the effective elastooptic coefficient is equal to $p_{ef}=p_{44}$=-0.078 for the direction of light wave [010]. Then the calculated value of AOFM amounts to $M_2$=49.95×10$^{-15}$s$^3$/kg, exceeding the corresponding coefficients for the fused quartz and KDP crystals. Besides, a small light absorption throughout the wide wavelength range, a high radiation threshold stability and a possibility for growing large samples with a high optical quality facilitate a practical choice of $\beta$-BaB$_2$O$_4$ crystals as an efficient AO material [3].

Constructing the indicative surface of the ultrasonic velocity for the $v_{13}$ transverse acoustic wave in *XY* plane demonstrates that the velocity achieves its lowest value (v=777m/s) when the *k* vector lies in *XY* plane and makes the angle of $\alpha$=57° with respect to *X* axis and the projections of the unit displacement vector are $X_x$=-0.142, $X_y$=0.115 and $X_z$=0.983 (Figure 1,b). The equation of the optical indicatrix in such a case

reads as

$$(B_1 + p_{11}e_1 + p_{12}e_2 + p_{13}e_3 + p_{14}e_4)X^2 +$$
$$+(B_1 + p_{12}e_1 + p_{11}e_2 + p_{13}e_3 - p_{14}e_4)Y^2 +$$
$$(B_3 + p_{31}e_1 + p_{31}e_2 + p_{33}e_3)Z^2 +$$
$$+2p_{41}(e_1 - e_2)ZY + 2p_{44}(e_4YZ + e_5XZ) = 1$$
(1)

One can neglect the terms that describe the optical indicatrix rotation as it has only a minor effect on the changes in the principal refractive indices and take into account that the incident optical wave is polarized parallel to $Z$ axis. Then the value of $p_{ef}$ should be determined by the change in the refractive index $n_3$, which is as follows:

$$\Delta n_3 \approx \frac{1}{2}n_3^3\{p_{31}e_1 + p_{31}e_2 + p_{33}e_3\} =$$
$$= \frac{1}{2}n_3^3\{-0.142p_{31} + 0.115p_{31} + 0.983p_{33}\}e$$
(2)

Taking the values of $p_{31}$=-0.112 and $p_{33}$=0.039, one can finally obtain $p_{ef}$=0.04 and $M_2$=12.32×10$^{-15}$s$^3$/kg. The decrease in $M_2$ in the specific latter case is related to the fact that the anisotropy of ultrasonic velocity is quite small (approximately 10%) and the other parameters, such as $p_{ef}$ and $n$, influence the $M_2$ value.

Table 1. AO parameters of β-BaB$_2$O$_4$ crystals ($\rho$=3840kg/m$^3$, $n_o$=1.6673 and $n_e$=1.5506 for $\lambda$=632.8nm).

| Acoustic wave | | p | \|p$_{eff}$\| | n | Light | | $M_2$, 10$^{-15}$ s$^3$/kg or possibility for matching the Bragg conditions |
|---|---|---|---|---|---|---|---|
| V, m/s | Propagation direction, Polarization | | | | Direction | Polarization | |
| 5410 | [100], [100] | p$_{11}$ | 0,195 | n$_e$ | [010] | [100] | not |
| | | p$_{31}$ | 0,112 | n$_o$ | | [001] | not |
| | | p$_{21}$ | 0,197 | n$_o$ | [001] | [010] | not |
| | | p$_{11}$ | 0,195 | n$_o$ | | [100] | not |
| 5805 | [010], [010] | p$_{11}$ | 0,195 | n$_e$ | [100] | [010] | not |
| | | p$_{42}$=-p$_{41}$ | 0,007 | n$_o$ | | [001], [010] | 0.0014 |
| | | p$_{31}$ | 0,112 | n$_o$ | | [001] | not |
| | | p$_{42}$=-p$_{41}$ | 0,007 | n$_e$ | | [010], [001] | 0.0009 |
| | | p$_{12}$ | 0,197 | n$_o$ | [001] | [100] | not |
| | | p$_{22}$=p$_{11}$ | 0,195 | n$_o$ | | [010] | not |
| 3500 | [001], [001] | p$_{23}$=p$_{13}$ | 0,059 | n$_e$ | [100] | [010] | not |
| | | p$_{33}$ | 0,039 | n$_o$ | | [001] | not |
| | | p$_{13}$ | 0,059 | n$_e$ | [010] | [100] | not |
| | | p$_{33}$ | 0,039 | n$_o$ | | [001] | not |
| 2900 | [100], [010] | p$_{56}$=p$_{41}$ | 0,007 | n$_o$ | [010] | [001], [100] | 0.01124 |
| | | p$_{41}$=p$_{56}$ | 0,007 | n$_e$ | | [100], [001] | 0.00726 |
| | | p$_{66}$ | 0.001 | n$_o$ | [001] | [010], [100] | not |
| | | p$_{66}$ | 0.001 | n$_o$ | | [100], [010] | not |
| 880 | [100], [001] | p$_{55}$=p$_{44}$ | 0,078 | n$_o$ | [010] | [001], [100] | 49.945 |
| | | p$_{55}$=p$_{44}$ | 0,078 | n$_e$ | | [100], [001] | 32.538 |
| | | p$_{65}$=p$_{14}$ | 0,005 | n$_o$ | [001] | [010], [100] | not |
| | | p$_{65}$=p$_{14}$ | 0,005 | n$_o$ | | [100], [010] | not |
| 940 | [010], [001] | p$_{24}$=-p$_{14}$ | 0,005 | n$_e$ | [100] | [010] | not |
| | | p$_{44}$ | 0,078 | n$_o$ | | [001], [010] | 40.979 |
| | | p$_{44}$ | 0,078 | n$_e$ | | [010], [001] | 26.452 |
| | | p$_{14}$ | 0,005 | n$_o$ | [001] | [100] | not |

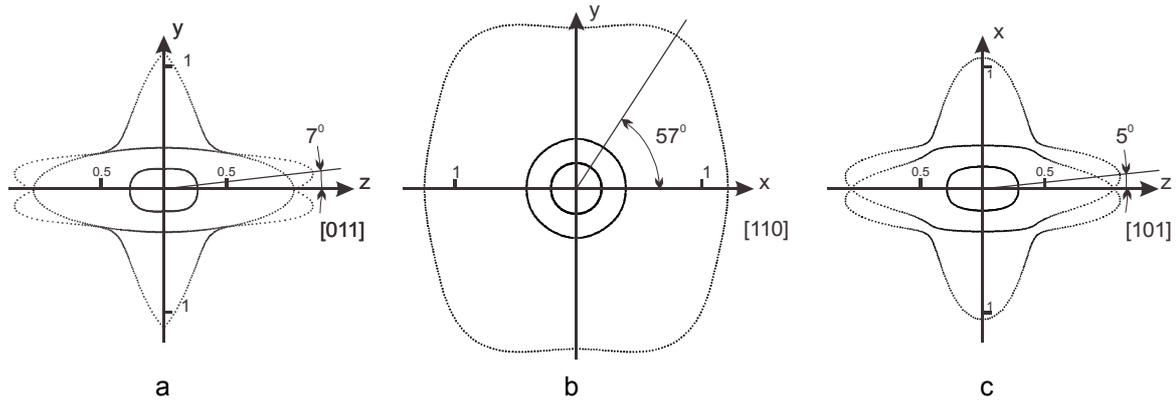

**Fig. 1.** Indicative surfaces of the inverse acoustic waves velocity for $\beta$-BaB$_2$O$_4$ crystals (in the units s×km$^{-1}$).

From other side, when the acoustic wave is propagate in $YZ$-plane with the direction of propagation which make the angle 7° with $Z$-axis and the projections of displacement vector $X_x=0$, $X_y=-0.99$, $X_z=0.125$ (Figure 1,a) the change of refractive index $n_3$ can be written as

$$\Delta n_3 \approx \frac{1}{2}n_3^3\{p_{31}e_2 + p_{33}e_3\} = \\ = \frac{1}{2}n_3^3\{-0.112 p_{31} + 0.039 p_{33}\}e \quad (3)$$

The value of effective photoelastic coefficient is equal $p_{ef}=0.12$, acoustic wave velocity – $v_{13}=817$m/s and AOFM – $M_2=95.35\times10^{-15}$s$^3$/kg. For the case of acoustic wave propagation in the $XZ$-plane (Figure 1, c) AOFM achieve only $M_2=2\times10^{-15}$s$^3$/kg.

Thus, the best geometry for the AO interaction in $\beta$-BaB$_2$O$_4$ crystals is that when the transverse acoustic wave $v_{13}$ with the velocity $v_{13}=817$m/s is propagated $YZ$-plane. AOFM in this case achieve value $M_2=95.35\times10^{-15}$s$^3$/kg.

## Acknowledgement

The authors are grateful to the Scientific and Technology Center of Ukraine for financial support of the present study (Project N1712).